\documentclass[a4paper]{article}

\usepackage{INTERSPEECH2022}

\title{CaTT-KWS: A Multi-stage Customized Keyword Spotting Framework based on Cascaded Transducer-Transformer}
\name{Zhanheng Yang$^{1,2*}$, Sining Sun$^{2*}$, Jin Li$^{2}$, Xiaoming Zhang$^{2}$, Xiong Wang$^{1}$,Long Ma$^{2}$, Lei Xie$^{1\dagger}$}
\address{
  $^1$Audio, Speech and Language Processing Group, School of Computer Science, Northwestern Polytechnical University, Xi'an, China \\
  $^2$Tencent Technology Co., Ltd, Beijing, China 
  \thanks{$^*$Equal contribution. $^\dagger$Lei Xie is the corresponding author.}}
\email{zhhyang@mail.nwpu.edu.cn,\{siningsun,hughjli,xiaomizhang,malonema\}@tencent.com,\\xwang@npu-aslp.org,
       lxie@nwpu.edu.cn }

\begin{document}

\maketitle
\begin{abstract}
Customized keyword spotting~(KWS) has great potential to be deployed on edge devices to achieve hands-free user experience. However, in real applications, false alarm (FA) would be a serious problem for spotting dozens or even hundreds of keywords, which drastically affects user experience. To solve this problem, in this paper, we leverage the recent advances in transducer and transformer based acoustic models and propose a new multi-stage customized KWS framework named Cascaded Transducer-Transformer KWS~(CaTT-KWS), which includes a transducer based keyword detector, a frame-level phone predictor based force alignment module and a transformer based decoder. Specifically, the streaming transducer module is used to spot keyword candidates in audio stream. Then force alignment is implemented using the phone posteriors predicted by the phone predictor to finish the first stage keyword verification and refine the time boundaries of keyword. Finally, the transformer decoder further verifies the triggered keyword. Our proposed CaTT-KWS framework reduces FA rate effectively without obviously hurting keyword recognition accuracy. Specifically, we can get impressively 0.13 FA per hour on a challenging dataset, with over 90\% relative reduction on FA comparing to the transducer based detection model, while keyword recognition accuracy only drops less than 2\%. 
\end{abstract}

\section{Introduction}

Keyword spotting (KWS) aims at detecting predefined keywords from consecutive audio stream. It has significant applications on edge devices to realize hands-free user experience. Besides detecting a single wake-up word (WuW), e.g., ``Alexa'' and ``Hi Google'' to initiate cloud-based speech interactions, on-device speech command recognition is also desired for command-and-control and privacy-sensitive applications. In these applications, an always-on KWS system runs locally on the resource-limited edge device, and thus needs to be small-footprint, prompt and resistant to false alarm~(FA) and false rejection~(FR). 
Recently, neural network approaches~\cite{0Convolutional,2014Small,wei2021end,jose2020accurate,berg2021keyword,qi2022exploiting,bae2018end} have been widely adopted. To ensure good performance, these approaches usually require training data for the specific sets of keywords, and the addition of a keyword to the current system requires a new round of data collection and model training. Moreover, with the increase of keywords for always-on applications, false alarms are notoriously hard to control. To trade-off, a typical solution is to cascade a wake-up word detection module with a windowed command recognition module. Once the wake-up word is triggered, another keyword spotting module is activated for a brief time slot only, for accepting user commands. In this paper, we aims to develop a new customized KWS framework with high accuracy as well as rare FA. More importantly, it is highly customizable to new keywords with no more effort on data collection and model training.


With natural streaming ability and its success in speech recognition~\cite{graves2012sequence,he2019streaming,sainath2020streaming,zhang2021tiny}, recurrent neural network transducer~(RNN-T) has recently been applied to KWS tasks as well~\cite{he2017streaming,liu2021rnn,tian2021improving,sharma2020adaptation}. Besides the streaming nature with low latency, transducer based acoustic model is flexible for open-vocabulary customized KWS as the modeling units can be sub-words, such as phonemes, which opens space for keyword customization. Some prior works on transducer based KWS aim to improve keyword accuracy by attention based biasing with predefined keyword transcripts~\cite{he2017streaming,liu2021rnn,tian2021improving,bluche2019predicting}. For example, He et al.~\cite{he2017streaming} proposed a technique to bias the search process towards a specific keyword using an attention mechanism, while Liu et al.~\cite{liu2021rnn} further improved the attention based biasing and added auxiliary loss function during model training. Recently, Tian et al.~\cite{tian2021improving} explored CTC joint training, stateless prediction network and various training data configuration strategies to avoid the over-fitting problem of transducer based KWS. 

Although transducer based KWS is flexible for keyword customization, mainly through biasing, there is still plenty of space for further improvement. First, previous studies mainly considered the wake-up scenario with a single keyword as target at runtime, while its ability in detecting multiple keywords simultaneously for speech command recognition scenario has not been explored. Second, transducer based methods have obtained impressively high wake-up rate, but FA is also a severe problem reported in the literature~\cite{liu2021rnn,tian2021improving}. This problem might be even severe for speech command recognition scenario, aiming at supporting dozens or hundreds of keywords at runtime.
Multi-stage strategy has been previously adopted to alleviate the problem~\cite{liu2021rnn,2020Multi, yang2021keyword,sigtia2021progressive}. In general, the first stage is a light-weight always-on keyword detector. Once a keyword candidate is detected, the corresponding audio segment is sent to the following stage(s) for further verification. In the multi-stage architecture, transducer is a great choice for the first stage due to its high recall on keywords. But how to design the following verification stages for FA reduction becomes crucial for the performance of the whole system. In~\cite{liu2021rnn}, which is also the most related work to our paper, a multi-level detection (MLD) method was proposed. The detection stage computes posterior sum confidence in a sliding window without regarding the phone order to detect keyword with a small computational cost. Subsequent verification stages compute edit distance probability confidence and approximate likelihood confidence as measurements to further verify the keyword. The MLD method is a statistics based verification method only relies on the transducer output.



In this paper, we propose a new neural network based multi-stage customized KWS framework named Cascaded Transducer-Transformer KWS~(CaTT-KWS), which includes a transducer based keyword detector, a frame-level phone predictor for force alignment and a transformer decoder. The three modules are shaped as a multi-task learning framework, where the encoder is shared across all the modules. Aiming at accurately spotting keyword candidates, the first stage adopts a tiny transducer~\cite{zhang2021tiny} cascaded with a WFST based decoding graph, using context independent (CI) phones as modeling units. Meanwhile, a rough time boundary of the triggered candidate will be generated for the following stages. This detection stage can be customized easily to accept user preferred keywords by simply modifying the search graph, without retraining the model. It is well-known that the output of streaming transducer has emission delay problem. Therefore, in the second stage,  Viterbi based force alignment is used to refine the keyword boundary and generate a likelihood score as confidence measure to decide if the audio segment generated in the detection stage includes a keyword or not. Lastly, a light-weight transformer decoder serves as the final verification stage, which accepts the encoder outputs corresponding to the triggered keyword and performs a beam search to finish the keyword verification. The proposed CaTT-KWS framework is verified on a challenging dataset, with impressively high keyword detection accuracy and low false alarms.



\section{Multi-stage Framework}
In this section, we introduce our proposed CaTT-KWS framework, including overview of the multi-task training procedure and detailed description on the design of the three stages -- the transducer detector, the force-alignment module and the transformer decoder.

\subsection{Multi-task learning procedure}

\begin{figure*}[!h]
	\vspace{-40pt}
	\centering 
	\includegraphics[width=1.0\linewidth]{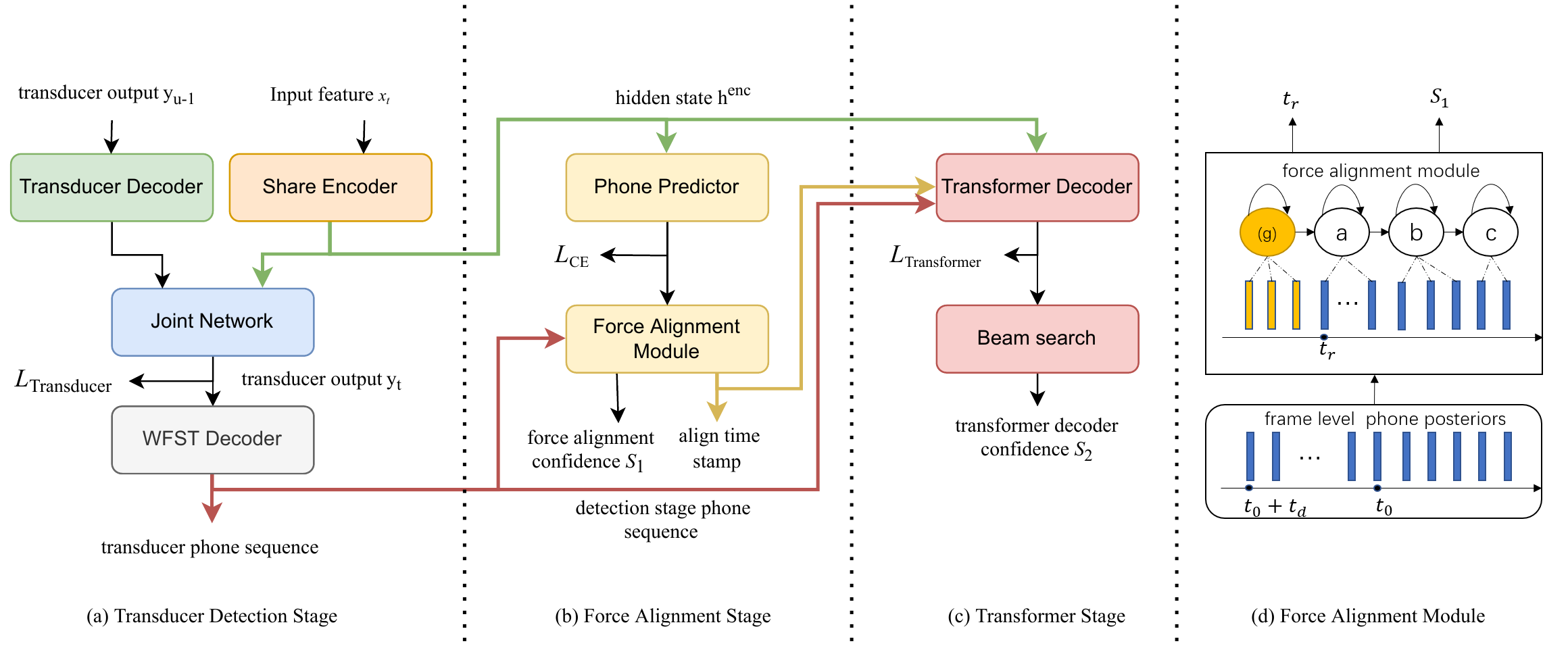}\vspace{-5pt}
	\caption{The proposed CaTT-KWS framework, which is composed of three stages -- transducer stage (a) for keyword detection and force alignment stage (b) as well as transformer stage (c) for keyword verification. Given the frame level phone posteriors generated by the phone predictor, the time stamp of the keyword candidate is further refined with more accurate start time stamp $t_r$ and likelihood score $S_1$, as shown in (d). 
	}\vspace{-18pt}
	\label{fig1}
\end{figure*}

We shape our multi-stage framework as a multi-task learning procedure, as shown in Fig.~\ref{fig1}. The overall framework can be regarded as cascaded transducer-transformer with a shared encoder. Furthermore, additional linear layer is added as a framewise phone predictor to predict phone label for each frame. On one hand, predicting frame-level phone label can accelerate and regularize the encoder training. On the other hand, during inference, once keywords are detected, the output of phone predictor will be used to obtain more accurate keyword boundary by force alignment. More details will be given in Section~\ref{ce}. The final loss function is a combination of transducer loss, transformer loss and frame-level cross-entropy (CE) loss, described as:
\begin{equation} \label{eq1}
	L = \alpha L_{\text{Transducer}} + \beta L_{\text{CE}} + \gamma L_{\text{Transformer}}
\end{equation}
where $\alpha$, $\beta$ and $\gamma$ are hyper-parameters. In this paper, we empirically set them to 1.0, 0.8 and 0.5 respectively. We performer multi-task learning during training and prepare corresponding transcript and frame-level phone alignment as label.


\subsection{Detection Stage: Tiny Transducer}
As shown in Fig.~\ref{fig1} (a), the transducer based detection stage consists of a tiny transducer based streaming acoustic model and a WFST based decoder. Note that this is similar to the hybrid framework~\cite{1990A,2017Compressed} which is widely adapted in KWS for keyword customization and restricting the search path. In this stage, the model tries to recall more cases on keywords, but it will inevitable result in a large number of false alarms.

Specifically, We adopt a tiny transducer proposed in~\cite{zhang2021tiny} which consists of a DFSMN-based~\cite{zhang2018deep}  encoder  and  a  casual  Conv1d based stateless predictor~\cite{ghodsi2020rnn}. The tiny transducer was specifically designed for edge device deployment with tricks for model compression and strategies for avoiding model over-fitting.


During inference, greedy search is used and only the posterior probabilities of non-blank outputs are fed into the WFST decoder, which is ``skip blank'' described in~\cite{zhang2021tiny}. 
Our decoding graph is composed by two separate WFSTs: lexicons ($L$) and predefined command sets as grammars ($G$). They are composed into the final $LG$ WFST for decoding, which can be presented as
\begin{equation}
	LG = \min(\det(L \circ G))~,
\end{equation}
where min and det represent minimize and determinize operations, respectively. Token passing algorithm is used to figure out the most likely triggered keyword and outputs a sketchy time boundaries of the keyword as well.


\subsection{Verification Stage: Force Alignment} \label{ce}
In our second stage, we aim to refine the keyword boundaries obtained in the previous detection stage for subsequent stage and reject parts of false alarms.

After the detection stage, we can get the keyword candidate's sketchy time boundary and the corresponding phone sequence during the Viterbi decoding. Hence in the verification stage, we aim to make use of the phone predictor to get a more accurate time stamp for the candidate location and further verify it, as shown in Fig.~\ref{fig1}(b).  Because of the well-known emission delay issue of streaming transducer, the boundary obtained from the detection stage is not accurate and this will lead to possible truncation of the candidate. To refine the boundary, as illustrated in Fig~\ref{fig1}(d), where $t_{0}$ is the original start point obtained from the transducer decoder, we push the start point empirically $t_d$ frame 
backward and feed the corresponding frames to the force alignment module with the phone sequence of the keyword candidate.




To implement force alignment, a simple linear WFST graph is constructed using the phone sequence of triggered candidate. Note that because we push the start point $t_d$ frames forward, which may introduce extra garbage segment, an extra symbol $(g)$  is inserted before the phone sequence to absorb the outputs that do not belong to the candidate keyword. Fig~\ref{fig1}(d) gives an keyword candidate example composed of phone sequence ``a, b, c" with extra garbage ``$(g)$". Finally we get the accurate start point $t_r$ of the candidate and the likelihood calculated during alignment is considered as the confidence score $S_1$ for this stage: 
\begin{equation} \label{eq3}
	S_1 = \frac{-\sum_{l\in f, t \in T}\log(p_{l}^{t})}{T}~,
\end{equation}
where $f$ denotes the framewise force alignment result and $p_{l}^{t}$ denotes the phone predictor posterior of label $l$ at $t^{th}$ frame. And $T$ is the accurate time stamp we get at this stage. Comparing $S_1$ with a pre-defined threshold $\tau$ can reject part of the false alarm examples effectively and positive examples with accurate boundaries are passed to the final verification stage.

\subsection{Verification Stage: Transformer Decoder}
In the final stage, we further verify the trigger command segment and make the final decision. After the above force alignment stage, more accurate boundaries of keyword candidate are obtained, with which the shared encoder's outputs $h^{enc}$ of the corresponding keyword candidate can be truncated out, which serves subsequently as the inputs of the transformer decoder, as shown in Fig.~\ref{fig1}(c). The transformer decoder accepts $h^{enc}$ and generates an $N$-best list autoregressively through beam search. Since we aim to confirm whether the segment is a real keyword or not, we do not need to search until $<$EOS$>$ appears. Therefore, we force to stop the beam search procedure after $M$ steps, where $M$ is the length of the phone sequence of the keyword candidate.  After decoding, we check if there is any output sequence that matches the phone sequence of the candidate. If none, it is rejected; otherwise, we compare score $S_2$ of the sequence to a preset threshold $\upsilon$. $S_2$ can be described as 
\begin{equation} \label{eq4}
	S_2 = -\sum_{l\in B}\log(p_{l})~,
\end{equation}
where $B$ denotes the triggered phone sequence in the beam. If $S_2$ is smaller than the threshold $\upsilon$, the candidate is eventually triggered as a keyword.


\vspace{-3pt}
\section{Experiments}

In this section, we introduce the corpus we used and describe the experimental setup about the detail of our model configuration and evaluation metrics. Experimental results and analysis are also presented at last. 

\subsection{Corpus}
Our models are trained on a set of 23,000-hour Mandarin ASR corpus which is collected from Tencent in-car speech assistant products. During training, the development set is randomly shuffled from the training set. We evaluated the accuracy of all models on a 6K utterances keyword set that cover 29 Mandarin commands (each has six or more phones), such as a command for starting GPS navigation that is consisted of phone sequence ``d a3 k ai1 d ao3 h ang2".
This 6K testing utterances are collected from both relatively `clean' highway driving conditions and noisy downtown driving conditions, resulting in the \textit{clean} set with 3K utterances and the \textit{noisy} set with another 3K utterances. For false alarm evaluation, we used an 84-hour audio set covering radio broadcast, chat and driving noise. All data we mentioned above are anonymous and hand-transcribed.

\subsection{Experimental Setup}
We used the 40-dim power-normalized cepstral coefficients (PNCC)~\cite{kim2016power} feature computed with a 25ms window and shifted every 10ms as input for all experiments. For the model configuration of our proposed model, the shared encoder consists of a convolution downsample layer to achieve a time reduction rate of 4 and a simple 1-layer 128-dim LSTM following a 6-layer 512-dim FSMN with a 320-dim linear projection. The left context of the FSMN module is 8 for all layer and right context is [2,2,1,2,2,1], respectively. The transducer decoder we used has a 1-D convolution layer with kernel size is 2. The joint network consists of a single feed-forward layer with 100 units following a hyperbolic tangent activation function. The output units include 210 context-independent (CI) phones and a blank symbol. The phone predictor only has one linear layer and a softmax layer, which maps the encoder outputs to phones. The transformer decoder includes several 512-dim self-attention blocks and we further explore the effect of the self-attention block's number in Section 3.4. The overall model size is about 3.8M, which is suitable for deploying on edge devices.

\subsection{Effect of Two Verification Stages}
\begin{figure}[t]
	\centering 
	\includegraphics[width=1.0\linewidth]{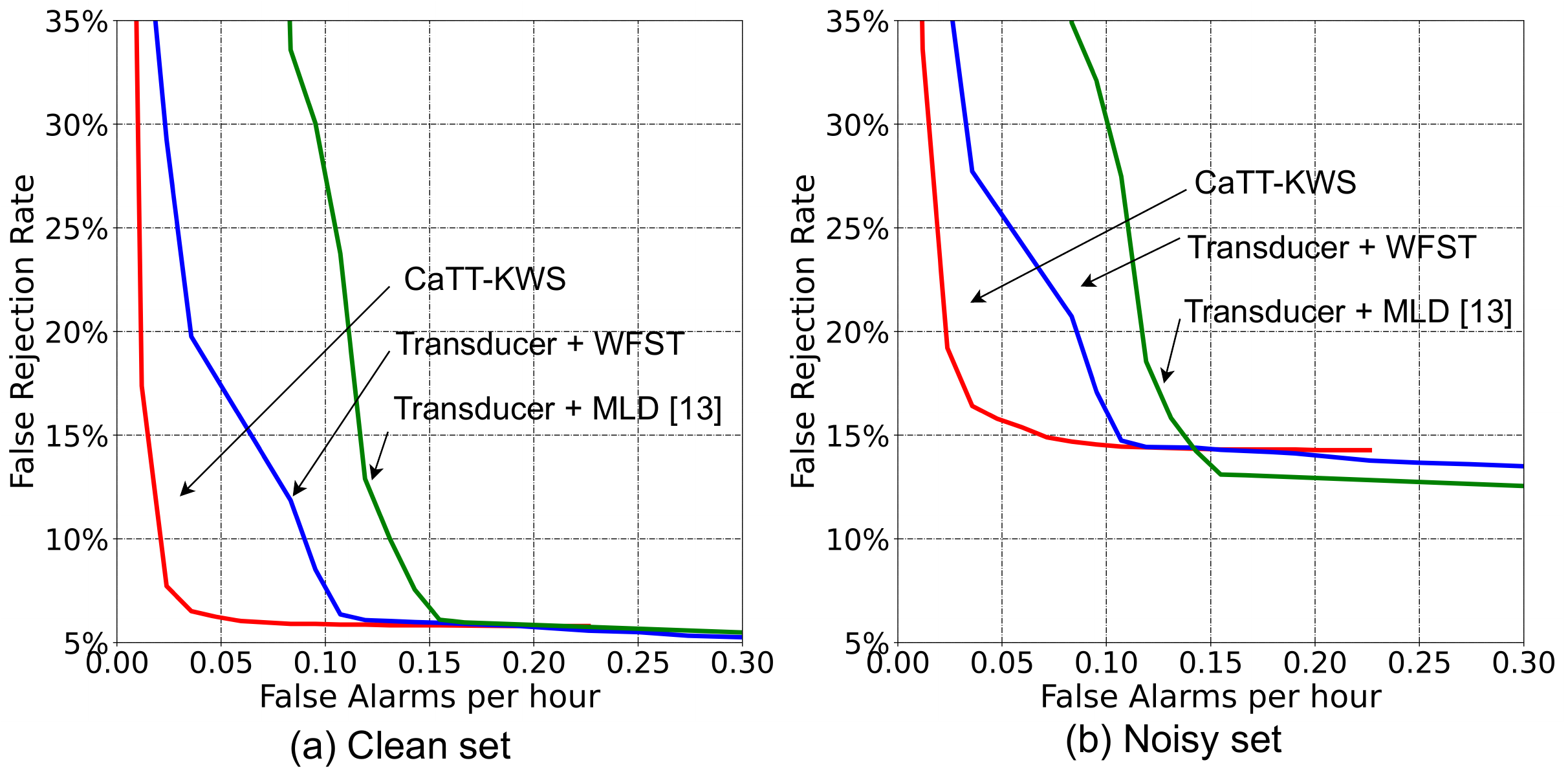}\vspace{-6pt}
	\caption{ROC curves for different systems.}\vspace{-15pt}
	\label{fig3}
\end{figure}

Table~\ref{tab:1} and Fig.~\ref{fig3} show the results of our proposed framework and MLD~\cite{liu2021rnn} method on clean and noisy evaluation sets. The force alignment stage threshold $\tau$ and transformer stage threshold $\upsilon$ are set as 1.5 and 5, respectively. And we set $t_d$ as 15 during the force alignment stage. The transformer decoder only has one attention block. From experiment S0 in Table~\ref{tab:1}, we find that in the detection stage, the accuracy rate of the tiny transducer can reach 96\% on the clean test set, which is pretty high. However, it also reaches a relative high FA at 1.47 per hour. By using our proposed force alignment and transformer stages, as shown in experiment S3, the FA per hour drops from 1.47 to 0.13 with little accuracy decay. Besides, we used the experiment setup suggested in~\cite{liu2021rnn} to evaluate the MLD method, and our proposed method can get a much lower FA rate with similar detection accuracy. 


\begin{table}[t]
\caption{Comparison on accuracy(\%) and FA number per hour among different framework. }\vspace{-5pt}
\label{tab:1}
\scalebox{0.8}{
\begin{tabular}{llclcc}
\hline
ID & Stage                  & \multicolumn{2}{c}{Acc (\%)}                          & \multicolumn{2}{c}{FA}                                       \\
   &                        & \multicolumn{1}{l}{Clean} & Noisy                     & \multicolumn{1}{l}{per hour} & \multicolumn{1}{l}{Gain (\%)} \\ \hline
S0 & Transducer + WFST      & 96                        & \multicolumn{1}{c}{87.65} & 1.47                         & -                             \\ \hline
S1 & + Transformer decoder  & 94.31                     & \multicolumn{1}{c}{85.89} & 0.23                         & 84.35                         \\ \hline
S2 & + Force alignment      & 95.48                     & \multicolumn{1}{c}{87.34} & 0.5                          & 65.98                         \\
S3 & ++ Transformer decoder & \multicolumn{1}{l}{94.20} & 85.69                     & 0.13                         & 91.15                         \\ \hline
S4 & Transducer + MLD       & \multicolumn{1}{l}{94.41} & 86.24                     & 0.27                         & -                             \\ \hline
\end{tabular}}\vspace{-10pt}
\end{table}

Fig~\ref{fig3} shows the receiver operating characteristic (ROC) curves of our proposed CaTT-KWS framework and the MLD method on the clean set and noisy set. During decoding, the threshold of transformer stage is fixed and threshold of force alignment stage changes dynamically. For the MLD method, only the threshold of last stage is changed to draw ROC curves. It is obvious that our proposed method outperforms the MLD method significantly when the number of FA (per hour) is less than 0.15, at which the FRR of MLD method worsens rapidly. Moreover, as shown in Table~\ref{tab:1}, accurate time boundaries of speech command can improve the performance of transformer decoder based verification comparing the result between the experiment of S1 (two stages framework) and S3 (three stages framework). In order to further analyze the result, we compare the average
timestamp offset compared with the ground truth of start point for the command phrase obtained by transducer detection stage and force alignment stage. As shown in Table~\ref{tab:2}, the timestamp generated from force alignment module is more accurate. And a accurate timestamp is very effective for reducing the number of FA



\begin{table}[t]
\caption{Comparison on start point error between transducer detection stage and force alignment stage. }\vspace{-5pt}
\centering
\label{tab:2}
\begin{tabular}{lcc}
\hline
Stage    & \multicolumn{1}{l}{Clean set (s)} & \multicolumn{1}{l}{Noisy set (s)} \\ \hline
Transducer detection  & 0.29                         & 0.44                         \\
Force alignment & 0.11                         & 0.23                         \\ \hline
\end{tabular}\vspace{-10pt}
\end{table}



\subsection{Impact of Transformer Decoder Size}
\begin{figure}[!h]
	\centering 
	\includegraphics[width=1.0\linewidth]{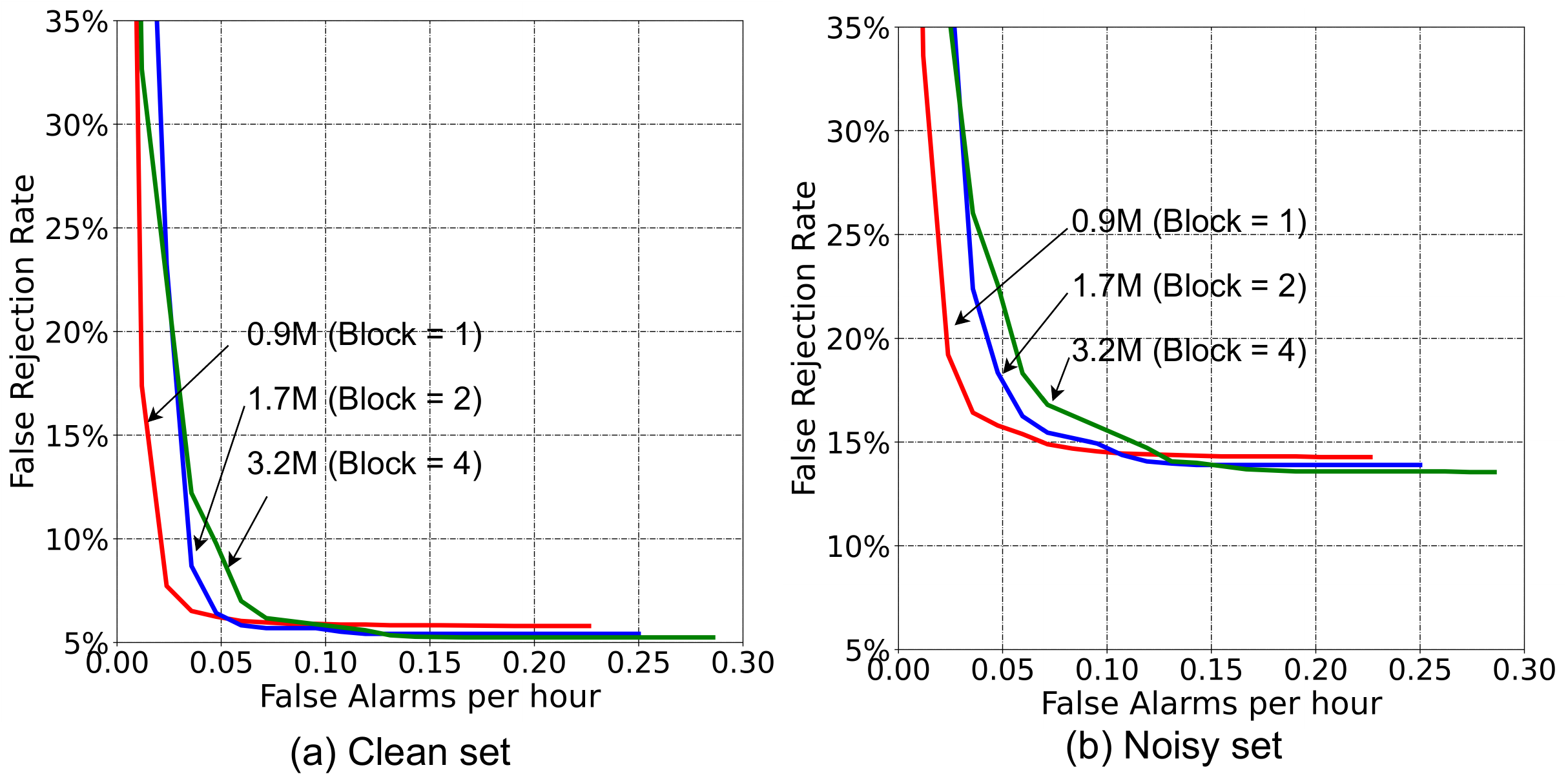}
	\caption{ROC curves for systems with different transformer decoder size.}\vspace{-8pt}
	\label{fig5}\vspace{-12pt}
\end{figure}

To explore how transformer decoder size can affect performance, we adjust the transformer decoder size by adding more self-attention blocks. The ROC results are reported in Fig.~\ref{fig5}. It denotes that adding transformer decoder size can not gain improvement on accuracy with similar FA levels while increasing the overall model size. The recognition performance even gets worse at low FA level because the overfitting of the model will be aggravated due to the use of more model parameters.

\subsection{Impact of Different Transformer Modeling Units}
\begin{figure}[!h]
	\centering 
	\includegraphics[width=1.0\linewidth]{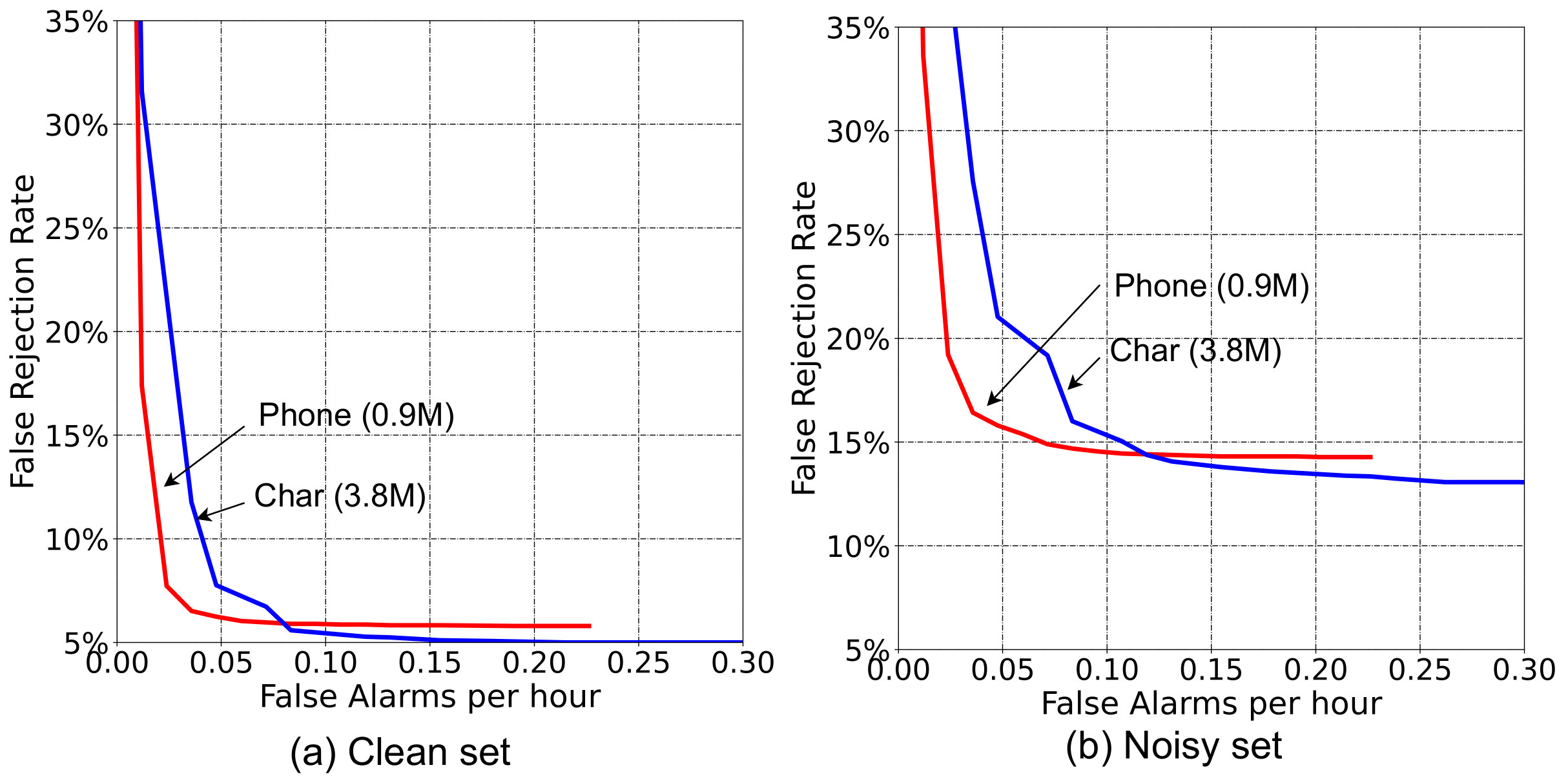}
	\caption{ROC curves for character- and phone- systems. The transformer decoder size is shown in parentheses.}\vspace{-8pt}
	\label{fig6}
\end{figure}

Considering the scalability of our model (e.g. customized commands), we choose CI phones as the model units of both transducer and transformer decoder in the experiments above. Here we further use 6,000 common Chinese characters as the modeling unit for the transformer decoder and make a comparison with the counterpart using CI phones. The ROC curves in Fig.~\ref{fig6} show that the phone model has a lower FR rate at lower FA conditions while the character model leads to an even lower FR rate at higher FA conditions. 
As a result, we can select appropriate modeling units according to different requirements, such as using modeling units with larger granularity like character in scenarios that require lower FRR. On the contrary, we can use units with smaller granularity like phone in scenarios that should pay more attention to FA.
By the way, because a larger number of model units result in bigger input and output layers, the size of transformer decoder of the character system becomes 4 times larger than that of the phone system. In order to effectively avoid the increase of model size, we can discard the unused input and output nodes according to the phones/characters covered in the predefined keyword set.

\section{Conclusions}
In this paper, we proposed a new KWS framework CaTT-KWS, aiming to suppress FA without obviously reducing command recognition accuracy. The model size of CaTT-KWS is about 3.8M, which is small enough for deploying on edge devices. Our proposed framework is shaped as the cascaded transducer-transformer architecture with the force alignment module.  
The force alignment module can also provide a more accurate time stamp. We evaluate our method on a sizable dataset, showing considerable relative reduction of over 90\% on FA,  which achieves about 0.13 FA per hour, while the command recognition accuracy only drops less than 2\%. And we further set up experiments to show the performance of our proposed CaTT-KWS framework, and explore the impact on transformer decoder size and model units as well.

\bibliographystyle{IEEEtran}

\bibliography{mybib}


\end{document}